\documentstyle[12pt]{article}
\begin{document}

\begin{titlepage}
\begin{flushright}
BA-98-48 \\
November 9, 1998 \\
\end{flushright}

\begin{center}
{\Large\bf  Neutrino Oscillations and Other \\
Key Issues in Supersymmetric \\ 
 SU(4)$_c \times $ SU(2)$_{\rm L} \times$ SU(2)$_{\rm R}$  
\footnote{Supported in part by  DOE under Grant No. DE-FG02-91ER40626
and by NATO, contract \\ 
~~~~~~~~~~~~number CRG-970149.}
}
\end{center}
\vspace{0.5cm}
\begin{center}
{\large Qaisar Shafi$^{a}$\footnote {E-mail address:
shafi@bartol.udel.edu} {}~and
{}~Zurab Tavartkiladze$^{b, c}$\footnote {E-mail address:
tavzur@axpfe1.fe.infn.it} }
\vspace{0.5cm}

$^a${\em Bartol Research Institute, University of Delaware,
Newark, DE 19716, USA \\

$^b$Istituto Nazionale di Fisica Nucleare, 
Sezione di Ferrara, 44100 Ferrara, Italy  \\

$^c$ Institute of Physics, Georgian Academy of Sciences,
380077 Tbilisi, Georgia}\\
\end{center}

\vspace{1.5cm}

\begin{abstract}

 We try to gain an understanding of the recent Superkamiokande
data on neutrino oscillations and several other 
important phenomenological
issues within the framework of  supersymmetric 
$SU(4)_c \times SU(2)_L \times SU(2)_R$
($\equiv G_{422}$). By supplementing $G_{422}$ with a 
$U(1)$-${\cal R}$ symmetry, we can  provide
an explanation of the magnitude $M_G$ ($\sim 10^{16}$~GeV) of the 
$G_{422}$- symmetry breaking scale, resolve
the MSSM $\mu $ problem, and understand why  proton decay 
has not been seen
($\tau_p \gg 10^{34}$~yr). 
The family dependent ${\cal R}$ - symmetry also helps provide
an explanation of the charged fermion mass hierarchies as well as the
magnitudes of the CKM matrix elements. 
Several additional heavy states in the mass range $10^4-10^7$~GeV are
predicted, and the MSSM parameter $\tan \beta $ turns out to be
of order unity.  The atmospheric neutrino problem is explained
through $\nu_{\mu}-\nu_{\tau }$ mixing with 
$\sin^2 2\theta_{\mu \tau }\simeq 1$.
The resolution of the solar 
neutrino puzzle is via the small angle MSW oscillations and necessarily 
requires a sterile neutrino 
$\nu_s$ which, thanks to the ${\cal R}$ - symmetry , 
has a tiny mass.

\end{abstract}

\end{titlepage}
\section{Introduction}

The Superkamiokande collaboration \cite{sup} may have presented 
the first `real'
experimental evidence of physics beyond the standard model through their
observations of the atmospheric and solar neutrino puzzles. These are
most simply resolved by invoking neutrino oscillations, 
thereby suggesting
that one or more of the known neutrinos has a mass greater 
than $\sim 10^{-1}-10^{-2}$~eV
or so. This value is far in excess of $\sim~ 10^{-5}$~eV that 
arises from the
`standard model' dimension five operators. Clearly, physics beyond the
standard model is called for, and supersymmetric unification is one 
good way to proceed.

Acceptable neutrino masses can be realized in several ways, but let us list
just two of them. One is the well-known see-saw mechanism \cite{seesaw} 
in which we invoke
the existence of a right-handed neutrino. In the second scheme 
\cite{tr} an $SU(2)_L$ 
scalar triplet naturally acquires a VEV of order $M_W^2/M$, where $M$ 
denotes the superheavy
mass of the triplet field. The neutrinos then acquire mass through their
coupling to the triplet VEV (Lepton number is not spontaneously broken
in either of these schemes). Of course, it is possible to consider models
in which both these mechanisms are simultaneously present.

In this paper we will focus on the 
$SU(4)_c \times SU(2)_L \times SU(2)_R$ ($\equiv G_{422}$) scheme
\cite{pati} which automatically introduces the right-handed 
neutrino. This scheme has 
a number of important features as pointed out in ref. \cite{king1}. 
First, in contrast to $SU(5)$ or $SO(10)$, it can provide 
a relatively straightforward
resolution of the MSSM $\mu $ problem. 
Second, once the hierarchy 
and $\mu $ 
problems are resolved it usually implies \cite{king1} an
essentially stable proton. Third, it contains a neat
mechanism for generating the observed baryon asymmetry via a lepton
asymmetry , with the right-handed neutrinos playing an essential role.
Last, but not least, the gauge symmetry $G_{422}$, 
again in contrast to
$SU(5)$ or $SO(10)$, can arise from fairly straightforward 
superstring constructions \cite{ant}.

The scheme presented here relies on the minimal `higgs' structure which gives
rise to MSSM at low energies. Remarkably, the gauge hierarchy 
and $\mu $ problems
are resolved within the minimal `higgs' framework
\footnote{The resolution
of the $\mu $ problem here will be different from the one 
given in \cite{king1}.}. An essential role in
achieving this is played by a (family dependent) 
$U(1)$~${\cal R}$-symmetry. This symmetry
also plays a crucial role in understanding the magnitude 
of the $G_{422}$ symmetry
breaking scale $M_G$ ($\sim 10^{16}$~GeV), which is 
expressed in terms of $M_P$
( reduced Planck scale $=2.4\cdot 10^{18}$~GeV) and $m_{3/2}$ 
($=$ gravitino mass
$\sim $~TeV). 
The ${\cal R}$-symmetry also implies an essentially stable 
proton ($\tau_p \sim 10^{60}$~yr) by strongly 
suppressing the dimension five operators from the colored triplets as
well as the nonrenormalizable Planck scale operators.

The ${\cal R}$ -symmetry also plays an essential role 
in realizing the observed
charged fermion mass hierarchies as well as the magnitudes of the CKM
matrix elements \cite{ram}. We are led to introduce a light sterile neutrino
in order to resolve the solar neutrino puzzle via the small angle
MSW solution \cite{bah}, while 
the atmospheric neutrino
puzzle is explained through large $\nu_{\mu }-\nu_{\tau }$ mixing.
We arrange for the sterile neutrino to be light by exploiting 
the ${\cal R}$- symmetry \cite{chun, shafi}.

The plan of the paper is as follows. In section 2  we consider the symmetry
breaking of $G_{422}$ as well as the origin of the GUT scale. 
Two important  
parameters make an appearance here. One is $\epsilon_{\cal R}  \sim 0.2$
(equal in magnitude to the Wolfenstein parameter 
$\lambda $), which is the ratio of 
the $U(1)$- ${\cal R}$
symmetry breaking scale and $M_P$, while the second parameter is
$\epsilon_G =
M_G/ M_P \sim 10^{-2}$ . In section 3  we discuss how the charged fermion
mass hierarchies and the magnitudes of the CKM matrix elements can be 
understood by exploiting the family dependent ${\cal R}$- symmetry. 
We are led to
introduce additional `matter' supermultiplets in the 
$(6,1,1)$ and $(1,2,2)$
representations of $G_{422}$ . It 
it turns out that the MSSM parameter
$\tan \beta $ has to be of order unity. 
In section 4 we discuss neutrino masses
and mixings in detail and show the need for a light sterile neutrino which,
thanks to the $U(1)$ -${\cal R}$- symmetry, is easily arranged.  
In section 5 we discuss
the stability of the proton and the important role 
played (yet again!) by the
${\cal R}$-symmetry. The conclusions are presented in section 6.

\section{$SU(4)_c\times SU(2)_L\times SU(2)_R$ Symmetry \\
Breaking and Origin of $M_G$}

The `higgs' sector
\footnote{We assume the existence of an unbroken $Z_2$ 
`matter' parity which
distinguishes the higgs and matter superfields.}
of the $SU(4)_c\times SU(2)_L\times SU(2)_R$ model
consists of the following superfields:

$$
H\sim (4, 1, 2)~,~~~~~~~~\bar H \sim (\overline{4}, 1, 2)~,
$$
\begin{equation}
h\sim (1, 2, 2)~,~~~~~~~~~D_6\sim (6, 1, 1)~,
\end{equation}
where

\begin{equation}
\begin{array}{cc}
(4, 1, 2)=~~ \\
\end{array}
\hspace{-6mm}\left(
\begin{array}{cccc}
\bar u^c_1& \bar u^c_2& \bar u^c_3&\overline{\nu }^c \\
\bar d^c_1& \bar d^c_2& \bar d^c_3&\bar e^c \end{array}
\right)~,~~~~
\begin{array}{cc}
(\overline{4}, 1, 2)=~~ \\
\end{array}
\hspace{-6mm}\left(
\begin{array}{cccc}
u^c_1& u^c_2& u^c_3&\nu^c \\
d^c_1& d^c_2& d^c_3&e^c \end{array}
\right)~,~~~~
\label {rep1}
\end{equation}

\begin{equation}
(1, 2, 2)=(h_u,~~h_d)~,~~~~~~~~(6, 1, 1)=(D^c,~~\bar D^c)~.
\label{rep2}
\end{equation}



 The $H$ and $\bar H$ fields provide the breaking 
of $G_{422}$ to 
$SU(3)_c\times SU(2)_L\times U(1)_Y$ after their components
$\overline{\nu }^c $ and $\nu^c $ develop nonzero VEVs. The states
$\bar u^c +u^c $ and $\bar e^c +e^c$ from $H+\bar H$ are `goldstones'
absorbed by the appropriate gauge fields, while $\bar d^c$ and $d^c$
form physical massive states through mixings with the corresponding
fragments from $D_6$. The superfield $h$ `unifies' 
the pair of MSSM electroweak
`higgs' doublets. For obtaining an all order doublet-triplet hierarchy
and the desirable pattern of symmetry breaking, we introduce an additional
singlet superfield $X$ and the crucial 
$U(1)$-${\cal R}$ symmetry. The latter
will be important also for understanding the hierarchies of fermion
masses and mixings. The ${\cal R}$ charges of the scalar superfields
and superpotential are presented in Table (\ref {t:scalar}).

\begin{table}
\caption{${\cal R}$ charges of the scalar
superfields
and the superpotential.}

\label{t:scalar}
$$\begin{array}{|c|c|c|c|c|c|c|}
\hline 

&W^{~} & H &\overline {H}  & D_6 &  h & X\\

\hline
&&&&&&\\
{\cal R}  &R &\frac{R}{10}+\frac{1}{2}R_X &\frac{R}{10}-\frac{1}{2}R_X &
\frac{4}{5}R-R_X & \frac{2}{5}R+\frac{1}{2}R_X 
&\frac{R}{24} \\ 
& &  & & & &\\
\hline
\end{array}$$
\end{table}

The scalar superpotential

\begin{equation}
W=M_P^3\left(\frac{\bar HH}{M_P^2}\right)^5
+M_P^3\left(\frac{X}{M_P}\right)^{24}+D_6HH+
\left(\frac{X}{M_P}\right)^2D_6\bar H\bar H~
\label{w}
\end{equation}
is the most general allowed under the ${\cal R}$ symmetry. 
The first two
terms in (\ref{w}), together with the quadratic soft terms

\begin{equation}  
V_{SSB}^m=m_{3/2}^2\left(|H|^2+|\bar H|^2 +|X|^2\right)~,
\label{ssb}
\end{equation}  
(which emerge in  $N=1$ SUGRA after SUSY breaking) lead to
the nonzero VEVs 
$\langle \overline{\nu }^c_H \rangle \equiv \langle H\rangle $ and
$\langle \nu^c_{\bar H} \rangle \equiv \langle \bar H \rangle $
with magnitudes:
\begin{equation}
\langle \bar H \rangle \sim \langle H\rangle
 \sim M_P\left(\frac{m_{3/2}}{M_P}\right)^{1/8}~,~~~~
\langle X \rangle \sim
M_P\left(\frac{m_{3/2}}{M_P} \right)^{1/22}~.
\label{vev}
\end{equation}
For $m_{3/2}= 10^3$~GeV and $M_P=2.4\cdot  10^{18}$~GeV, 
we have:
$$
\epsilon_G \equiv \frac{\langle H \rangle }{M_P}\sim 
\left(\frac{m_{3/2}}{M_P} \right)^{1/8}\sim 10^{-2}~,~~~~
$$
\begin{equation}
\epsilon_{\cal R} \equiv \frac{\langle X\rangle }{M_P}
\sim
\left(\frac{m_{3/2}}{M_P} \right)^{1/22}\sim 0.2~.
\label{eps}
\end{equation}
%
%
%
We observe that the ${\cal R}$ symmetry 
(broken at scale $\langle X \rangle $) helps  generate 
the GUT scale with magnitude 
$M_{G}\simeq \epsilon_G M_P \sim 10^{16}~$GeV.

The last two terms in (\ref{w}) are responsible
for the generation of masses of the colored triplet fragments.
Substituting the nonzero VEVs of the scalars in these two terms
the masses of the decoupled states are given by:

\begin{equation}
m_{T_1}\simeq M_G\simeq M_P\epsilon_G~,~~~~~~~~~~
m_{T_2}\simeq M_G\epsilon^2_{\cal R}
\simeq M_P\epsilon_G \epsilon_{\cal R}^2~.
\label{mt}
\end{equation}

Note, that the electroweak higgs doublets are massless to all  orders
in the unbroken SUSY limit.
One source for obtaining the desirable value of 
the MSSM $\mu $ term can be the mechanism
suggested in ref. \cite{giu}. It generates the $\mu $
term through the K\"ahler potential, and can be applied also in our  
scheme. Consider the coupling between the electroweak doublets and
the hidden sector field $Z$:

\begin{equation}
\Delta K_h=\frac{Z}{M_P}h^2~.
\label{kal}
\end{equation}
For $\langle F_Z \rangle \sim m_{3/2}M_P$,
one obtains $\mu \sim m_{3/2}$.

In summary, the $U(1)$ ${\cal R}$ symmetry leads to the desirable
gauge hierarchy and is crucial for the generation of the GUT scale
\footnote{For an explanation of the origin of GUT scale and all order
natural gauge hierarchy in $SU(3)^3$ and flipped $SU(6)$
models, see \cite{gia1} and \cite{shafi} respectively.}.

\section{Horizontal $U(1)$ ~${\cal R}$ Symmetry: \\
Charged Fermion Masses and Mixings}

In the simplest approach to $G_{422}$ the `matter' 
sector consists of the superfields
$F_{\alpha }\sim (4, 2, 1)_{\alpha }$,
$\bar F_{\alpha }\sim (\overline{4}, 1, 2)_{\alpha }$ ($\alpha $ is
a family index), where
$F_{\alpha } \supset (q,~l)_{\alpha },~
\bar F_{\alpha } \supset (u^c,~ d^c,~e^c)_{\alpha }$,
and the masses of the quark-lepton families are
generated through the coupling
${\cal A}_{\alpha \beta }F_{\alpha }\bar F_{\beta }h$.
If the family dependent ${\cal A}_{\alpha \beta }$ couplings
are taken as field independent Yukawa constants, one obtains
the  unacceptable
asymptotic relations $\hat{Y}_u=\hat{Y}_d=\hat{Y}_e $ and no CKM
mixings emerge. However, if appropriate entries of ${\cal A}$ are
$\left( \frac{\bar HH}{M^2}\right) $ ($M$ is some cut-off) dependent operators
then, due to the many possible  convolutions of
$G_{422}$ group indices, the unwanted mass
relations are avoided and physical mixings can emerge. The mixing
angles and values of Yukawa couplings depend on the physical cut-off
$M$, and taking it $M_P$ (which could be considered as a natural
cut-off of the theory), one observes that all the mixing 
angles are much
smaller than the corresponding measured 
values.
Consequently, for obtaining reasonable values of the CKM matrix elements
the cut-off $M$ should be taken to be somewhat lower 
\cite{king}-\cite{alla3}, assuming
that the corresponding operators are obtained after decoupling of
some states whose masses lie between the scales $M_{G}$ and $M_P$.
Refs. \cite{alla1, alla2} present the classification of corresponding
operators and some needed $G_{422}$
representations of the decoupled states.

Our approach here is rather different. The cut-off parameter of
all nonrenormalizable operators which we consider will be the Planck
mass. For obtaining the desirable hierarchies of Yukawa couplings and
mixings we introduce three additional pairs of `matter' superfields:

\begin{equation}
g_{\alpha}\sim (6, 1, 1)_{\alpha }~,~~~~~
f_{\alpha}\sim (1, 2, 2)_{\alpha }~.
\label{gfrep}
\end{equation}
These representations turn out to be the  most  economical
ones for building a phenomenologically acceptable
`matter' sector of our model \footnote{It is worth noting that these
additional states, together with $F_{\alpha }$, $\bar F_{\alpha }$
and singlet states ${\cal N}_2, {\cal N}_3$ and $\nu_s$
 which we will introduce
later for accommodating the Superkamiokande data, constitute the
(three) ${27}_{\alpha }$-plets of $E_6$.}.

We will consider the ${\cal R}$ symmetry as `horizontal' 
and prescribe distinct $R$ charges to fermions from 
different families in order to obtain a natural 
explanation of the hierarchies
of Yukawa couplings and CKM matrix elements.
The transformation properties of the various `matter' superfields
are presented in Table (\ref{t:fer}) .

\begin{table}
\caption{$R$ charges of the `matter' superfields
under the ${\cal R}$  symmetry.
}
\label{t:fer}
$$\begin{array}{|c|c|c|c|c|c|}
\hline
 &F_1&F_2&F_3 & \overline {F}_1& \overline {F}_2 \\
\hline
~ &~ &~ &~ &~ &~ \\
{\cal R}&\frac{3}{10}R-\frac{7}{2}R_X &\frac{3}{10}R-\frac{5}{2}R_X &
 \frac{3}{10}R-\frac{1}{2}R_X & \frac{3}{10}R-3R_X
 &\frac{3}{10}R-R_X \\
~&~&~&~&~&~\\
\hline
\hline
&\overline {F}_3 &g_1 &g_2 &g_3 &f_{\alpha }\\
\hline
~&~ &~ &~ &~ &~  \\
{\cal R}&\frac{3}{10}R  &\frac{1}{5}R-\frac{5}{2}R_X   &
 \frac{1}{5}R-\frac{1}{2}R_X  & \frac{1}{5}R-\frac{1}{2}R_X
 &\frac{2}{5}R-(9-\alpha )R_X  \\
&~ &~ &~ &~ &~  \\
\hline
\end{array}$$
\end{table}

The 
$F_{\alpha }\bar F_{\beta }h$ type couplings are schematically
written as

\begin{equation}
\begin{array}{ccc}
 & {\begin{array}{ccc}
\bar F_1   &\,\,~~~~~\bar F_2~~~  &\,\,~~\bar F_3~~~~
\end{array}}\\ \vspace{6mm}
\begin{array}{c}
F_1\\ F_2 \\ F_3

\end{array}\!\!\!\!\! &{\left(\begin{array}{ccc}
\,\, \left(\frac{X}{M_P}\right)^6 &\,\,\left(\frac{X}{M_P}\right)^4
&\,\,\left(\frac{X}{M_P}\right)^3
\\
\,\, \left(\frac{X}{M_P}\right)^5 &
\,\,\left(\frac{X}{M_P}\right)^3  &\,\,
\left(\frac{X}{M_P}\right)^2    \\

\,\,\left(\frac{X}{M_P}\right)^3&\,\,\frac{X}{M_P}
  &\,\,1

 \end{array}~\right)\cdot h}
\end{array}  \!\!~,~~~
\label{FF}
\end{equation}
and upon diagonalization yields the up quark Yukawa matrix:

\begin{equation}
\hat{Y}_u^D={\rm Diag}\left(\epsilon_{\cal R}^6,~~
\epsilon_{\cal R}^3,~~1 \right)~,
\label{upd}
\end{equation}
where the top quark Yukawa coupling
\begin{equation}
\lambda_t \sim  1~,
\label{top}
\end{equation}
and
\begin{equation}
\lambda_u : \lambda_c :  \lambda_t \sim
\epsilon_{\cal R}^6 : \epsilon_{\cal R}^3 :1~.
\label{lamdau}
\end{equation}

The couplings involving the $g$ states  read:

\begin{equation}
\begin{array}{ccc}
 & {\begin{array}{ccc}
g_1~~  &\,\,~~~~g_2~~~~  &\,\,~~g_3~~~
\end{array}}\\ \vspace{6mm}
\begin{array}{c}
F_1\\ F_2 \\ F_3
 
\end{array}\!\!\!\!\! &{\left(\begin{array}{ccc}
\,\, \left(\frac{X}{M_P}\right)^5 &\,\,\left(\frac{X}{M_P}\right)^3 
&\,\,\left(\frac{X}{M_P}\right)^3
\\ 
\,\, \left(\frac{X}{M_P}\right)^4 &
\,\,\left(\frac{X}{M_P}\right)^2  &\,\,
\left(\frac{X}{M_P}\right)^2    \\

\,\,\left(\frac{X}{M_P}\right)^2&\,\,1
  &\,\,1     
\end{array}~\right)\frac{Hh}{M_P}} 
\end{array}  \!\!~,~~~
\begin{array}{ccc}
 & {\begin{array}{ccc} 
\bar F_1~~~ &\,\,~~\bar F_2~~~ &\,\,~~\bar F_3~~~~~~~~~~~
\end{array}}\\ \vspace{2mm}
\begin{array}{c}
g_1\\ g_2 \\ g_3
 
\end{array}\!\!\!\!\! &{\left(\begin{array}{ccc}
\,\, \left(\frac{X}{M_P}\right)^6 &
\,\,\left(\frac{X}{M_P}\right)^4 &
\,\,\left(\frac{X}{M_P}\right)^3
\\ 
\,\,  \left(\frac{X}{M_P}\right)^4 &\,\, 
\left(\frac{X}{M_P}\right)^2&
\,\,\frac{X}{M_P}    \\

\,\, \left(\frac{X}{M_P}\right)^4 &
\,\,\left(\frac{X}{M_P}\right)^2 &\,\,\frac{X}{M_P}  
 \end{array}\right)\left(\frac{\bar HH}{M_P^2}\right)^2 \bar H } 
\end{array}  
\label{FggF}
\end{equation}

\begin{equation}
\begin{array}{ccc}
 & {\begin{array}{ccc} 
g_1 &\,\,
~~~~~g_2~~~~  &\,\,~~g_3~~~~~~~~~~~~~
\end{array}}\\ \vspace{2mm}
\begin{array}{c}
g_1 \\ g_2   \\ g_3 
 
\end{array}\!\!\!\!\! &{\left(\begin{array}{ccc}
\,\, \left(\frac{X}{M_P}\right)^5
 &\,\,\left(\frac{X}{M_P}\right)^3 &\,\,\left(\frac{X}{M_P}\right)^3   
\\ 
\,\, \left(\frac{X}{M_P}\right)^3 &\,\,\frac{X}{M_P} 
&\,\,\frac{X}{M_P}
\\
\,\, \left(\frac{X}{M_P}\right)^3 &
\,\,\frac{X}{M_P} &\,\,\frac{X}{M_P}

 \end{array}\right)\left(\frac{\bar HH}{M_P^2}\right)^3M_P } 
\end{array} ~.~~
\label{gg}
\end{equation}
Without loss of generality we consider a basis in which
the couplings (\ref{FF}) and (\ref{gg}) are diagonal,
and the matrix
relevant for down-quark masses will be:

\begin{equation}
\hat{M_d}=\begin{array}{cc}
 & {\begin{array}{cc}
~~D^c_g~~~  &\,\,~~d^c~~~~
\end{array}}\\ \vspace{2mm}
\begin{array}{c}
d \\ \bar D^c_g
\end{array}\!\!\!\!\! &{\left(\begin{array}{cc}
\,\, \hat{m}_d' &\,\,~~\hat{Y}_u^Dh_d
\\
 \,\, \hat{M}_g &\,\,~~\hat{M}_{g\bar F}

\end{array}\right)}~~,
\end{array}   ~~
\label{down1}
\end{equation}
where, taking into account (\ref{FggF}) and (\ref{gg}),

\begin{equation}
\begin{array}{ccc}
\hat{m}_d'=~~ \\
\end{array}
\hspace{-6mm}\left(
\begin{array}{ccc}
\epsilon_{\cal R}^5& \epsilon_{\cal R}^3 & \epsilon_{\cal R}^3 \\
\epsilon_{\cal R}^4& \epsilon_{\cal R}^2& \epsilon_{\cal R}^2 \\
\epsilon_{\cal R}^2& 1 & 1 \end{array}
\right)\epsilon_G h_d ~,~~
\begin{array}{ccc}
\hat{M}_{g\bar F}=~~ \\
\end{array}
\hspace{-6mm}\left(
\begin{array}{ccc}
b_{11}\epsilon_{\cal R}^5& b_{12}\epsilon_{\cal R}^3 & 
b_{13}\epsilon_{\cal R}^2 \\
b_{21}\epsilon_{\cal R}^3& b_{22}\epsilon_{\cal R}& b_{23} \\
b_{31}\epsilon_{\cal R}^3& b_{32}\epsilon_{\cal R} & b_{33} \end{array}
\right)M_P\epsilon_G^5\epsilon_{\cal R} ~,
\label {mats1}
\end{equation}

\begin{equation}
\hat{M}_g={\rm Diag}\left(\epsilon_{\cal R}^4,~1,~1\right)
M_P\epsilon_G^6\epsilon_{\cal R}
\label{gd}
\end{equation}
and $\hat{Y}_u^D$ is given in (\ref{upd}). For $b_{12}, b_{13}< 1/3$
the states $\bar D^c_g$ and $d^c$ can integrated out, 
and so for the down quark mass matrix we obtain:

\begin{equation}
\begin{array}{ccc}
 & {\begin{array}{ccc}
~d^{c'}_1& \,\,~~~d^{c'}_2  & \,\,~~d^{c'}_3~~~~~
\end{array}}\\ \vspace{2mm}
\hat{m}_d=\hat{m}_d'-Y_u^D\hat{M}_{g\bar F}^{-1}M_gh_d\simeq 
\begin{array}{c}
q_1\\ q_2 \\q_3 
 \end{array}\!\!\!\!\! &{\left(\begin{array}{ccc}
\,\,\epsilon_{\cal R}^5  &\,\,~~\epsilon_{\cal R}^3 &
\,\,~~\epsilon_{\cal R}^3 
\\ 
\,\,\epsilon_{\cal R}^4   &\,\,~~\epsilon_{\cal R}^2  &
\,\,~~\epsilon_{\cal R}^2
 \\
\,\, \epsilon_{\cal R}^2 &\,\,~~ 1  &\,\,~~1 
\end{array}\right) \epsilon_G h_d}~. 
\end{array}  \!\!  ~~~~~
\label{down2}
\end{equation}
From (\ref{down2}), the down quark Yukawa couplings are:

$$
\lambda_b\sim \epsilon_G ~(\sim 10^{-2}) ~,~~~
$$
\begin{equation}
\lambda_d :\lambda_s :\lambda_b \sim 
\epsilon_{\cal R}^5:\epsilon_{\cal R}^2 :1~.
\label{lambdad}
\end{equation}
These values lead us to conclude that the 
MSSM parameter $\tan \beta $ 
($\equiv \langle h_u \rangle /\langle h_d\rangle $) 
is of order unity. For the CKM matrix elements we obtain:
\begin{equation}
V_{us}\sim \epsilon_{\cal R}~,~~~V_{ub}\sim \epsilon_{\cal R}^3~,~~~
V_{cb}\sim \epsilon_{\cal R}^2~,
\label{ckm}
\end{equation}
which are in good agreement with the observations!

Turning now to the lepton sector, the fields $f_{\alpha }$ 
are crucial for obtaining values
of the charged lepton Yukawa couplings that are consistent with the
$\tan \beta \sim $ unity regime. The couplings
containing these fields are:

\begin{equation}
\begin{array}{ccc}
 & {\begin{array}{ccc}
f_1~~~~~ &\,\,~~f_2~~~~~ &\,\,~~f_3~~~~~~~
\end{array}}\\ \vspace{2mm}
\begin{array}{c}
F_1\\ F_2 \\ F_3

\end{array}\!\!\!\!\! &{\left(\begin{array}{ccc}
\,\, \left(\frac{X}{M_P}\right)^{12} &
\,\,\left(\frac{X}{M_P}\right)^{11} &
\,\,\left(\frac{X}{M_P}\right)^{10}
\\
\,\,  \left(\frac{X}{M_P}\right)^{11} &\,\,
\left(\frac{X}{M_P}\right)^{10}&
\,\,\left(\frac{X}{M_P}\right)^{9}  \\
\,\,\left(\frac{X}{M_P}\right)^9 &
\,\,\left(\frac{X}{M_P}\right)^8 &\,\,\left(\frac{X}{M_P}\right)^{7}
\end{array}\right)\frac{\bar HH}{M_P^2}\bar H }
\end{array}  \!\!~,~~
\label{Ff}
\end{equation}

\begin{equation}
\begin{array}{ccc}
 & {\begin{array}{ccc}
~f_1~~~~~ &\,\,
~~f_2~~~~  &\,\,~~f_3~~~~~~
\end{array}}\\ \vspace{2mm}
\begin{array}{c}
f_1 \\ f_2   \\ f_3

\end{array}\!\!\!\!\! &{\left(\begin{array}{ccc}
\,\, \left(\frac{X}{M_P}\right)^{16}
 &\,\,\left(\frac{X}{M_P}\right)^{15} &\,\,
\left(\frac{X}{M_P}\right)^{14}
\\
\,\, \left(\frac{X}{M_P}\right)^{15}&\,\,\left(\frac{X}{M_P}\right)^{14}
&\,\,\left(\frac{X}{M_P}\right)^{13}
\\
\,\,\left(\frac{X}{M_P}\right)^{14} &\left(\frac{X}{M_P}\right)^{13}
&\,\,  \left(\frac{X}{M_P}\right)^{12}
\end{array}\right)\frac{\bar HH}{M_P}}
\end{array} ~,~~
\label{ff}
\end{equation}
and, in the basis where (\ref{ff}) is taken diagonal, we will have:

\begin{equation}
\hat{M_e}=\begin{array}{cc}
 & {\begin{array}{cc} 
l_f~~~~~ &\,\,~l~~~~
\end{array}}\\ \vspace{2mm}
\begin{array}{c}
e^c \\ \bar e_f
\end{array}\!\!\!\!\! &{\left(\begin{array}{cc}
\,\, 0 &\,\,~~\hat{Y}_u^Dh_d
\\
 \,\, \hat{M}_f &\,\,~~\hat{M}_{fF}   

\end{array}\right)}~~,
\end{array}   ~~
\label{lep1}
\end{equation}
with

\begin{equation}
\begin{array}{ccc}
\hat{M}_{fF}=~~ \\
\end{array}
\hspace{-6mm}\left(
\begin{array}{ccc}
c_{11}\epsilon_{\cal R}^5& c_{12}\epsilon_{\cal R}^4 & 
c_{13}\epsilon_{\cal R}^2 \\
c_{21}\epsilon_{\cal R}^4& c_{22}\epsilon_{\cal R}^3& 
c_{23}\epsilon_{\cal R} \\
c_{31}\epsilon_{\cal R}^3& c_{32}\epsilon_{\cal R}^2 & c_{33} \end{array}
\right)M_P\epsilon_G^3\epsilon_{\cal R}^7 ~,~~~~
\hat{M}_f={\rm Diag }\left(\epsilon_{\cal R}^4,~\epsilon_{\cal R}^2,~1 \right)
M_P\epsilon_G^2\epsilon_{\cal R}^{12}
\label {mats2}
\end{equation}
For $c_{13}, c_{23}<1/5$, the states $l-\bar e_f$ can be
integrated out and for the charged lepton matrix one obtains:

\begin{equation}
\hat{m}_e=Y_u^D\hat{M}_{fF}^{-1}\hat{M}_fh_d~.
\label{lep2}
\end{equation}
More explicitly,

\begin{equation}
\begin{array}{ccc}
 & {\begin{array}{ccc} 
\hspace{-5mm}~l_{f_1}'~ & \,\,~l_{f_2}'~~  & \,\,~l_{f_2}'~~~
\end{array}}\\ \vspace{2mm}
\hat{m}_e= \begin{array}{c}
e^c_1\\ e^c_2 \\e^c_3
 \end{array}\!\!\!\!\! &{\left(\begin{array}{ccc}
\,\,\epsilon_{\cal R}^5  &\,\,~~\epsilon_{\cal R}^4 &
\,\,~~\epsilon_{\cal R}^3 
\\ 
\,\,\epsilon_{\cal }^3   &\,\,~~\epsilon_{\cal R}^2  &
\,\,~~\epsilon_{\cal R}
 \\
\,\, \epsilon_{\cal R}^2 &\,\,~~ \epsilon_{\cal R} &\,\,~~1 
\end{array}\right) \frac{\epsilon_{\cal R}^5}{\epsilon_G } h_d}~. 
\end{array}  \!\!  ~~~~~
\label{lep3}
\end{equation}
Diagonalizing (\ref{lep3}) we find the charged lepton Yukawa
couplings to be:

$$
\lambda_{\tau }\sim \frac{\epsilon_{\cal R}^5}
{\epsilon_G }\sim 10^{-2} (\sim \lambda_b )~,
$$
\begin{equation}
\lambda_e :\lambda_{\mu } :\lambda_{\tau } \sim
\epsilon_{\cal R}^5:\epsilon_{\cal R}^2 :1~.
\label{lamdae}
\end{equation}
Thus, with the help of $g+f$ states and imposing a horizontal
${\cal R}$ symmetry, we can  indeed realize the desirable 
hierarchies of
Yukawa couplings and magnitudes of CKM mixing angles, as given
by (\ref{top}), (\ref{lamdau}), (\ref{lambdad}), (\ref{lamdae}) and
(\ref{ckm}) respectively.

From (\ref{mats1})  and (\ref{mats2}) one finds that the 
new triplet and doublet `matter' states lie below
the GUT scale:

$$
m_{t_1}\simeq M_P\epsilon_G^5\epsilon_{\cal R}^6~,~~~
m_{t_2}\simeq M_P\epsilon_G^5\epsilon_{\cal R}^2~,~~~
m_{t_3}\simeq M_P\epsilon_G^5\epsilon_{\cal R}~,
$$
\begin{equation}
m_{d_1}\simeq M_P\epsilon_G^3\epsilon_{\cal R}^{12}~,~~~
m_{d_2}\simeq M_P\epsilon_G^3\epsilon_{\cal R}^{10}~,~~~
m_{d_3}\simeq M_P\epsilon_G^3\epsilon_{\cal R}^7~.
\label{mdt}
\end{equation}
These masses vary from around $10^4-10^7$~GeV.
Also, from  (\ref{mt})  the mass of one triplet pair from
the `higgs' sector is below $M_G$.
The `observed' unification at $M_G$ \cite{lan} of the three  gauge 
coupling constants in the one
loop approximation will still hold if the masses of these 
additional states satisfy the condition:

\begin{equation}
\frac{m_{T_2}}{M_G}\simeq
\frac{m_{d_1}m_{d_2}m_{d_3}}{ m_{t_1}m_{t_2}m_{t_3}}~.
\label{cond}
\end{equation}
Substituting  the masses of the corresponding
triplet and doublet fragments from  (\ref{mt}) and (\ref{mdt}),
we can see that (\ref{cond}) is indeed
satisfied. 

\section{Neutrino Oscillations}

From the couplings (\ref{FF}), (\ref{Ff}) and (\ref{ff}) the
matrix relevant for neutrino `Dirac' masses turns out to be:

\begin{equation}
\hat{M}_{\nu }'=\begin{array}{cc}
 & {\begin{array}{cc}
\nu_f~~~~~ &\,\,~\nu ~~~~
\end{array}}\\ \vspace{2mm}
\begin{array}{c}
\nu^c \\ \overline {\nu }_f
\end{array}\!\!\!\!\! &{\left(\begin{array}{cc}
\,\, 0 &\,\,~~\hat{Y}_u^Dh_u
\\
 \,\, \hat{M}_f &\,\,~~\hat{M}_{fF}

\end{array}\right)}~~.
\end{array}   ~~
\label{matnu1}
\end{equation}
Integrating out the $\overline {\nu }_f-\nu $ states yields
for the `light' sector:

\begin{equation}
\hat{m}_D'=Y_u^D\hat{M}_{fF}^{-1}\hat{M}_fh_u~.
\label{nu}
\end{equation}
The matrix (\ref{nu}) coincides with and has the same 
orientation in family space
as the charged lepton mass matrix (\ref{lep2}), and by proper unitary
transformations, can be diagonalized and expressed as:

\begin{equation}
\hat{m}_D'= {\rm Diag}\left(
\lambda_e~,~ \lambda_{\mu }~,~\lambda_{\tau }\right)h_u.
\label{nud}
\end{equation}
These `Dirac' masses will be `suppressed' by the see-saw mechanism
\cite{seesaw}. The ${\cal R}$ symmetry allowed couplings:

\begin{equation}
\begin{array}{ccc}
 & {\begin{array}{ccc}
\hspace{-6mm}\bar F_1~  &\,\,
~~~~~\bar F_2~~
&~~~~\bar F_3~~~~
\end{array}}\\
\begin{array}{c}
\bar F_1\\ \bar F_2 \\ \bar F_3

\end{array}\!\!\!\!\! &{\left(\begin{array}{ccc}
\,\, \left( \frac{X}{M_P}\right)^5 &\,\,\left( \frac{X}{M_P}\right)^3 &
\,\,\left( \frac{X}{M_P}\right)^2
\\
\,\, \left( \frac{X}{M_P}\right)^3 &
\,\,\frac{X}{M_P}  &
\,\,1   \\

\,\,\left( \frac{X}{M_P}\right)^2&\,\,1 &
\,\,0

 \end{array}\right)\frac{\bar HH}{M_P^3}HH }
\end{array}  \!\!~
\label{maj}
\end{equation}
generate the `Majorana' matrix for $\nu^c $ states which, 
in the basis in which $\hat{m}_D'$ is diagonal (see (\ref{nud})),
takes the form:

\begin{equation}
\begin{array}{ccc}
\hat{M}_R'=~~ \\
\end{array}
\hspace{-6mm}\left(
\begin{array}{ccc}
\epsilon_{\cal R}^5& \epsilon_{\cal R}^3 & \epsilon_{\cal R}^2 \\
\epsilon_{\cal R}^3& \epsilon_{\cal R} & 1 \\
\epsilon_{\cal R}^2& 1 & \epsilon_{\cal R}  \end{array}
\right)M_P\epsilon_G^4 ~.~~~~
\label {matmaj}
\end{equation}
Finally, for the neutrino masses we have:
\begin{equation}
\hat{m}_{\nu }'=\hat{m}_D'(\hat{M}_R')^{-1}\hat{m}_D'~,
\label{nu1}
\end{equation}
from which, taking into account (\ref{lamdae}), 

\begin{equation}
m'_{\nu_{\tau} }\sim
\frac{\epsilon_{\cal R}\lambda_{\tau }^2}{M_P\epsilon_G^4}h_u^2~,~~~
m'_{\nu_{\mu }}\sim \epsilon_{\cal R}^2m'_{\nu{\tau }}~,~~
m'_{\nu_e} \sim \epsilon_{\cal R}^4m'_{\nu_{\tau }}~.
\label{nu1mass}
\end{equation}
For $\lambda_{\tau }\sim 10^{-2}$, $h_u=174$~GeV 
we find

\begin{equation}
m_{\nu_{\tau }}'\simeq 2.5\cdot 10^{-2}~{\rm eV}~,~~~~~
m_{\nu_{\mu }}'\simeq  10^{-3}~{\rm eV}~,~~~~~
m_{\nu_e}'\simeq 4\cdot 10^{-5}~{\rm eV}~.
\label{nu2mass}
\end{equation}
These values of neutrino masses and the corresponding mixing 
angles 
($\theta_{12}'\sim \theta_{23}'\sim \epsilon_{\cal R}~,
\theta_{13}'\sim \epsilon_{\cal R}^2$)
are clearly inconsistent with the atmospheric and 
solar neutrino data 
\cite{sup}. We have to suitably modify 
our scheme and the 
mechanism that we propose is relatively simple,
although its realization requires additional singlet 
states ${\cal N}_{2,3}$ and $\nu_s$ (see footnote 7). 
Transformation properties of ${\cal N}_{2,3}$ superfields
under ${\cal R}$  given in Table (\ref{t:singlets}).

\begin{table}
\caption{$R$ charges of the singlet states ${\cal N}_2$, ${\cal N}_3$.
}
\label{t:singlets}
$$\begin{array}{|c|c|c|}
\hline
 &{\cal N}_2 &{\cal N}_3 \\
\hline
& &  \\
{\cal R}& \frac{1}{5}R-\frac{13}{2}R_X &
\frac{1}{5}R+\frac{13}{2}R_X   \\
& &  \\
\hline
\end{array}$$
\end{table}

We begin with the atmospheric neutrino puzzle. Noting that the 
physical `light' left-handed neutrinos $\nu_i$  
reside mainly in the $f_i$ (see (\ref{matnu1})), the relevant
terms for the
$\nu_{\mu }-\nu_{\tau }$ system are:

\begin{equation}
\begin{array}{cc}
 & {\begin{array}{cc}
~~{\cal N}_2~&\,\,~~{\cal N}_3~
\end{array}}\\ \vspace{2mm}
\begin{array}{c}
f_2\\ f_3 \\ 
 
\end{array}\!\!\!\!\! &{\left(\begin{array}{ccc}
\,\, \left(\frac{X}{M_P}\right)^{13} &
\,\,1
\\ 
\,\,  \left(\frac{X}{M_P}\right)^{12} &\,\,0     
\end{array}\right)h } 
\end{array}  \!\!~,~~~~~~~~~~~
\begin{array}{cc}
 & {\begin{array}{cc} 
{\cal N}_2~~ &\,\,
{\cal N}_3~~~~~~~~~~~  
\end{array}}\\ \vspace{2mm}
\begin{array}{c}
{\cal N}_2 \\ {\cal N}_3
 
\end{array}\!\!\!\!\! &{\left(\begin{array}{ccc}
\,\, \left(\frac{X}{M_P}\right)^{13}
 &\,\,1    
\\ 
\,\, 1
&\,\,0
\end{array}\right)\left(\frac{\bar HH}{M_P^2}\right)^3M_P} 
\end{array} ~,~~
\label{fNNN}
\end{equation}
from which we get:

\begin{equation}
\begin{array}{cc}
{\hat{m}_D= \left(\begin{array}{cc}
\,\, \epsilon_{\cal R}^{13} &
\,\,1
\\
\,\,  \epsilon_{\cal R}^{12} &\,\,0     
\end{array}\right)h_u } 
\end{array}  \!\!~,~~~~~~~~
\begin{array}{cc}
{\hat{M}_{{\cal N}_{2,3}}= \left(\begin{array}{ccc}
\,\, \epsilon_{\cal R}^{13}
 &\,\,1    
\\ 
\,\, 1
&\,\,0
\end{array}\right)M_P\epsilon_G^6} 
\end{array} ~~,~
\label{mats3}
\end{equation}

\begin{equation}
\begin{array}{cc}

\hat{m}_{\nu }=\hat{m}_D\hat{M}_{{\cal N}_{2,3}}^{-1}\hat{m}_D^T=
\!\!\!\!\! &{\left(\begin{array}{cc}
\,\,\epsilon_{\cal R}  &\,\,~~1 
\\ 
\,\, 1 &\,\,~~0 
\end{array}\right) \frac{\epsilon_{\cal R}^{12}h_u^2}{M_P\epsilon_G^6 }}~. 
\end{array}  \!\!  ~~~~~
\label{nu2}
\end{equation}

Thus, the neutrino mass matrix involving the second and third
generations has the quasi degenerate form, which provides
the large mixing
as well as the needed mass squared difference:

$$
m_{\nu_2}\simeq m_{\nu_3}\equiv m\sim 
\frac{\epsilon_{\cal R}^{12}h_u^2}{M_P\epsilon_G^6 }\simeq
5\cdot 10^{-2}~{\rm eV}~,
$$
$$
\Delta m_{23}^2=2m^2\epsilon_{\cal R}\simeq 10^{-3}~{\rm eV}^2~,
$$
\begin{equation}
\sin^2 2\theta_{\mu \tau } \simeq 1~.
\label{atm}
\end{equation}
For obtaining this picture the form of
$\nu_{\mu }-\nu_{\tau }$ mass matrix in (\ref{nu2}) 
is crucial
\footnote{The $\nu_{\mu }-\nu_{\tau }$ mass matrix of our model
closely resembles the one given in ref. \cite{barg}.}, and we must 
ensure the absence of terms which can spoil the large mixing.
The contribution from the elements of $\hat{m}_{\nu }'$
(see (\ref{nu1mass})-(\ref{nu2mass})) are negligible, 
and also inclusion of the
first generation does not change the picture.

The solar neutrino puzzle in our scheme can be explained through
the small angle MSW oscillations.
For this we have to invoke a new sterile state $\nu_s$ with 
${\cal R}$ charge equal to $R_{\nu_s }=-19R_X/2$.  
The relevant superpotential couplings are:

\begin{equation}
W_{\nu_e s}=\left(\frac{\bar HH}{M_P^2} \right)
\left(\frac{X}{M_P}\right)^{17}\nu_sf_1h+
M_P
\left(\frac{X}{M_P}\right)^{43}\nu_s^2~,
\label{wsol}
\end{equation}
which give

\begin{equation}
\begin{array}{cc}
 & {\begin{array}{cc} 
~\nu_e &\,\,
~~\nu_s~~
\end{array}}\\ \vspace{2mm}
\hat{m}_{\nu_e \nu_s}=
\begin{array}{c}
\nu_e \\ \nu_s    
 \end{array}\!\!\!\!\! &{\left(\begin{array}{cc}
\,\, 0
 &\,\,m_{\nu_e \nu_s}    
\\ 
\,\, m_{\nu_e \nu_s}
&\,\,m_{\nu_s }
\end{array}\right)}
\end{array} ~~,~
\label{nusol}
\end{equation}
with the corresponding entries in the range:
$$
m_{\nu_e \nu_s}\sim 
\epsilon_G^2\epsilon_{\cal R}^{17}h_u\sim 
\left(2.3\cdot 10^{-5}-1.2\cdot 10^{-4}\right)~{\rm eV}~,
$$
\begin{equation}
m_{\nu_s }\sim M_P\epsilon_{\cal R}^{43}
\sim \left(2\cdot 10^{-3}-1.3\cdot 10^{-1}\right)~{\rm eV},
\label{nust}
\end{equation}
which for $m_{\nu_s }\simeq  10^{-3}$~eV, 
$m_{\nu_e \nu_s}\simeq 5\cdot 10^{-5}$~eV gives  
the oscillation parameters:

$$
\Delta m_{\nu_e \nu_s}^2\simeq m_{\nu_s}^2
\sim  10^{-6}~{\rm eV}^2
$$
\begin{equation}
\sin^2 2\theta_{es}\sim 10^{-2} ~,
\label{sol}
\end{equation}
which correspond to the small angle MSW oscillations of $\nu_e$ 
into the sterile state $\nu_s$ \cite{bah}. 
We repeat that in our model the small 
masses $m_{\nu_e \nu_s}$ and $m_{\nu_s}$ 
(see (\ref{nusol}), (\ref{nust}) ) 
are guaranteed by the ${\cal R}$ symmetry.

Considering all generations together and taking into account
the relevant couplings, 
we have checked that after decoupling of the heavy states,
the results obtained for the atmospheric and solar neutrino 
oscillation parameters are
unchanged.

\section{Proton Decay}

For studying nucleon decay in our model we begin with 
the color triplet mass matrix  which arises from the `higgs' sector.
From  (\ref{w}) we have:

\begin{equation}
\begin{array}{cc}
 & {\begin{array}{cc}
~d^c_{\bar H} &\,\,
~~D^c_D~~
\end{array}}\\ \vspace{2mm}
\hat{M}_T=
\begin{array}{c}
\bar d^c_H \\ \bar D^c_D
 \end{array}\!\!\!\!\! &{\left(\begin{array}{cc}
\,\, m_{3/2}
 &\,\,M_P\epsilon_G
\\
\,\, M_P\epsilon_G \epsilon_{\cal R}^2
&\,\,0
\end{array}\right)}
\end{array} ~.~~
\label{trmat}
\end{equation}

Baryon and lepton number violating $d=5$ operators, obtained
after integrating out these triplet states, will be proportional 
to the elements of the matrix $\hat{M}_T^{-1}$ given by:

$$
\left(\hat{M}_T^{-1} \right)_{11}\simeq 0~,~~~
\left(\hat{M}_T^{-1} \right)_{22}\simeq
\frac{m_{3/2}}{M_P^2\epsilon_G^2 \epsilon_{\cal R}^2 }~,
$$
\begin{equation}
\left(\hat{M}_T^{-1} \right)_{12}\simeq
\frac{1}{M_P\epsilon_G \epsilon_{\cal R}^2 }~,
~~~~
\left(\hat{M}_T^{-1} \right)_{21}\simeq 
\frac{1}{M_P \epsilon_G }~.
\label{trmatel}
\end{equation}
From  (\ref{trmat}) and (\ref{trmatel}) we see that nucleon
decay can occur if the superfields $D_6$ and $H, \bar H$
simultaneously have  couplings with the relevant matter superfields.
It is easy to verify that the couplings

$$
FFD_6~,~~~~~~~~~~~FfD_6H~,
$$
\begin{equation}
\bar F\bar F D_6~,~~~~~~~~~~~\bar FgD_6H~,
\label{matterD}
\end{equation}
involving the $D_6$ field are forbidden by ${\cal R}$ symmetry. 
This fact ensures that dimension five colored triplet induced
nucleon decay is absent in our model. 

Similarly, non-renormalizable operators suppressed by the
Planck scale such as:

$$
{\cal O}_L^{(1)}=\frac{1}{M_P}FFFF~,~~~~~~
{\cal O}_L^{(2)}=\frac{1}{M_P^2}FFFfH~,~~~~~~
$$
\begin{equation}
{\cal O}_R^{(1)}=\frac{1}{M_P}\bar F\bar F\bar F\bar F~,~~~~~~
{\cal O}_R^{(2)}=\frac{1}{M_P^2}\bar F\bar F\bar FgH~,
\label{d5ops}
\end{equation}
(which can lead to $d=5$ operators  $qqql$ and $u^cu^cd^ce^c$ 
respectively), are also eliminated by the ${\cal R}$ symmetry.

Note that it is possible
that the zero entry in (\ref{trmat}) is replaced by a term of order
$m_{3/2}$ due to contributions from the K\"ahler potential
and the hidden sector (which can also give rise to 
couplings of the type in (\ref{matterD}), (\ref{d5ops})). 
These effects would induce proton decay with lifetime
$\tau_p \sim 10^{60}$~yr.

We have also checked that dimension five operators involving the
sterile neutrino superfield $\nu_s$, such as $u^cd^cd^c\nu_s$,
are also strongly suppressed. They also imply a proton lifetime
$\sim 10^{60}$~yr.

We therefore conclude that the proton is essentially stable
in the $SU(4)_c\times SU(2)_L\times SU(2)_R$ scheme discussed 
here. Its lifetime is estimated to be $\tau_p\sim 10^{60}$~yr
\footnote{Let us note here that using a $U(1)$- ${\cal R}$ 
symmetry to eliminate the Planck scale $d=5$ operators is also
discussed in ref. \cite{shafi}.}. 

\section{Conclusions}

In this paper we have attempted  a unified treatment of 
several important phenomenological problems within the 
framework of supersymmetric
$SU(4)_c\times  SU(2)_L \times  SU(2)_R$ ($\equiv G_{422}$). 
It is quite remarkable that by supplementing $G_{422}$ with 
a single family dependent $U(1)$- ${\cal R}$ symmetry and $Z_2$
matter parity, one can obtain an understanding of a wide 
ranging set of phenomena. For instance, one can explain the 
origin of the symmetry scale $M_G$ of $G_{422}$, understand 
why the MSSM $\mu $ term is of order a TeV or so
rather than $M_P$, provide an estimate of the MSSM parameter 
$\tan \beta $ (order unity), understand why proton decay has 
not been (and will not be!) seen, and gain an understanding 
of fermion mass hierarchies and the magnitude of the
CKM matrix elements.The model predicts the existence of new 
`heavy' (mass $\sim 10^4-10^7$GeV) particles. Lastly, 
and perhaps most significantly,
the model can also accomodate the recent Superkamiokande data.
The small angle $\nu_e - \nu_s$ MSW 
oscillations resolve the solar neutrino puzzle, 
while $\nu_{\mu }-\nu_{\tau }$ oscillations 
(with $\sin^2 2\theta_{\mu \tau }\simeq 1$) 
are responsible for the atmospheric neutrino anomaly. 
The sterile
neutrino $\nu_s$ is kept light by the ${\cal R}$ symmetry.

\vspace{.5cm}
{\bf {Acknowledgements}}

We thank Dr. John Bahcall for bringing to our attention the fact
that the large angle $\nu_e-\nu_s$ vacuum oscillations are excluded at 
a high confidence level by the solar neutrino data.


\end{document}